\documentclass[apj,twocolumn]{emulateapj}
\usepackage{natbib}
\usepackage{float}
\usepackage{graphics,rotating,multirow,bm,color,amsmath,amssymb,hyperref}
\usepackage{epsfig}

\hypersetup{
    colorlinks=true,
    linkcolor=green,
    citecolor=blue,
}

\begin{document}

\title{Varying Alpha from $N$-Body Simulations}
\author{Baojiu~Li$^{1,2}$, David~F.~Mota$^3$ and John~D.~Barrow$^2$}
\affiliation{$^1$DAMTP, Centre for Mathematical Sciences, University of Cambridge, Wilberforce Road, Cambridge CB3 0WA, UK} 
\affiliation{$^2$Kavli Institute for Cosmology Cambridge, Madingley Road, Cambridge CB3 0HA, UK}
\affiliation{$^3$Institute of Theoretical Astrophysics, University of Oslo, 0315 Oslo, Norway}
\email[Email address: ]{b.li@damtp.cam.ac.uk}
\email[Email address: ]{d.f.mota@astro.uio.no}
\email[Email address: ]{j.d.barrow@damtp.cam.ac.uk}

\begin{abstract}
We have studied the Bekenstein-Sandvik-Barrow-Magueijo (BSBM) model for the
spatial and temporal variations of the fine structure constant, $\alpha $,
with the aid of full $N$-body simulations which explicitly and
self-consistently solve for the scalar field driving the $\alpha $-evolution
. We focus on the scalar field (or equivalently $\alpha $) inside the dark
matter halos and find that the profile of the scalar field is essentially
independent of the BSBM model parameter. This means that given the density
profile of an isolated halo and the background value of the scalar field, we
can accurately determine the scalar field perturbation in that halo. We also
derive an analytic expression for the scalar-field perturbation using the
Navarro-Frenk-White halo profile, and show that it agrees well with
numerical results, at least for isolated halos; for non-isolated halos this
prediction differs from numerical result by a (nearly) constant offset which
depends on the environment of the halo.
\end{abstract}

\maketitle


\section{Introduction}

\label{sect:intro}

During the past ten years there has been continual interest in the possible
time and space variation of fundamental constants of Nature. This interest
was stimulated by observations of quasar absorption spectra that are
consistent with a slow increase of the fine structure \textquotedblleft
constant\textquotedblright , $\alpha $, over cosmological time scale %
\citep{Webb:1999, Webb:2001}. Experimental and observational efforts to
constrain the level of any possible time variation in fundamental constants
have a history that pre-dates modern theories about how they might vary (for
overviews see \cite{revs1, revs2}, \cite{revs3} and \cite{revs4}). Until
quite recently all the observational studies found no evidence for any
variations. However, high-quality data from a number of astronomical
observations have provided evidence that at least two of these constants,
the fine structure constant: $\alpha =e^{2}/$ $\hbar c$, and the
electron-proton mass ratio: $\mu =m_{e}/m_{p}$ might have varied slightly
over cosmological time. Using a data set of 128 \texttt{KECK-HIRES} quasar
absorption systems at redshifts $0.5<z<3$, and a new many-multiplet (MM)
analysis of the line separations between many pairs of atomic species
possessing relativistic corrections to their fine structure, \cite%
{Webb:1999, Webb:2001} found the observed absorption spectra to be
consistent with a shift in the value of fine structure constant, $\alpha $,
between those redshifts and the present day, with $\Delta \alpha /\alpha
\equiv \alpha (z)-\alpha (0)/\alpha (0)=-0.57\pm 0.10\times 10^{-5}$. A
smaller study of 23 \texttt{VLT-UVES} absorption systems between $0.4\leq
z\leq 2.3$ by \cite{Chand:2004} and \cite{Srianand:2004} initially found $%
\Delta \alpha /\alpha =-0.6\pm 0.6\times 10^{-6}$ by using an approximate
version of the full MM technique. However, the reanalysis of the same data
set by \cite{murphyrev, murphyrev2} using the full and unbiased MM method
increased the uncertainties and suggested a revised figure of $\Delta \alpha
/\alpha =-0.64\pm 0.36\times 10^{-5}$ for the same data. These
investigations relied on the statistical gain from large samples of quasar
absorption spectra. Most recently, this observational programme has been
extended to both hemispheres of the sky using \texttt{KECK} and \texttt{VLT}
samples of 153 absorption systems by \cite{Webb:2010} and finds evidence
consistent with an increase in $\alpha $ the northern sky but consistent
with a slow decrease in $\alpha $ with time in the south. When combined
these overlapping data sets are well fitted by a dipole with $\ \Delta
\alpha /\alpha _{0}=(1.10\pm 0.25)\times 10^{-6}r\cos \theta $, at
measurement position \b{r} (relative to Earth at \b{r} = 0) where $\theta $
is the angle between the measurement and the axis of the dipole. These
observations suggest that we should develop an understanding of the spatial
as well as the temporal consequences of varying constants.

By contrast to these statistical searches for varying $\alpha $, probes of
the electron-proton mass ration can use single objects effectively. \cite%
{reinhold} have found a $3.5\sigma $ indication of a variation in the
electron-proton mass ratio $\mu $ $\equiv m_{e}/m_{pr}$ over the last $12\,%
\mathrm{Gyrs}$: $\Delta \mu /\mu =(-24.4\pm 5.9)\times 10^{-6}$ from $H_{2}$
absorption in a different object at $z=2.8$. However, \cite{murphmass} have
exploited the $\mu $ sensitivity of ammonia inversion transitions %
\citep{flam} compared to rotational transitions of CO, HCN, and HCO$^{+}$ in
the direction of the quasar B0218+357 at $z=0.68466$ to yield a result that
is consistent with no variation in $\mu $ when systematic errors are more
fully accounted for: $\Delta \mu /\mu =(+0.74\pm 0.47_{\mathrm{stat}}\pm
0.76_{\mathrm{system}})\times 10^{-6}$, corresponding to a time variation of 
$\dot{\mu}/\mu =(-1.2\pm 0.8_{\mathrm{stat}}\pm 1.2_{\mathrm{system}})\times
10^{-16}$yr$^{-1\text{ }}$ in the best-fit $\Lambda$CDM cosmology.

Any variation of $\alpha $ and $\mu$ today could also be constrained by
direct laboratory searches. These are performed by comparing clocks based on
different atomic frequency standards over a period of months or years. Until
very recently, the most stringent constraints on the temporal variation in $%
\alpha $ arose by combining measurements of the frequencies of Sr %
\citep{blatt}, Hg+ \citep{fortier}, Yb+ \citep{peiknew}, and H %
\citep{fischer} relative to Caesium: $\dot{\alpha}/\alpha =(-3.3\pm
3.0)\times 10^{-16}\, \mathrm{yr}^{-1}$. \cite{cingoz} also recently
reported a less stringent limit of $\dot{\alpha}/{\alpha }=-(2.7\pm
2.6)\times 10^{-15}\, \mathrm{yr}^{-1}$; however, if the systematics can be
fully understood, an ultimate sensitivity of $10^{-18}\,\mathrm{yr}^{-1}$ is
possible with their method \cite{nguyen}. If a linear variation in $\alpha $
is assumed with cosmic time then the \cite{Webb:1999, Webb:2001} quasar
measurements equate to $\dot{\alpha}/{\alpha }=(6.4\pm 1.4)\times 10^{-16}\, 
\mathrm{yr}^{-1}$. If the variation is due to a light scalar field described
by a theory like that of Bekenstein, Sandvik, Barrow and Magueijo (\texttt{%
BSBM} from here on) \citep{bek, Sandvik:2001}, then the rate of change in
the constants is exponentially damped during the recent
dark-energy-dominated era of accelerated expansion, and one typically
predicts $\dot{\alpha}/\alpha =1.1\pm 0.3\times 10^{-16}\,\mathrm{yr}^{-1}$
from the Murphy \emph{et al.} data, which is not ruled out by the atomic
clock constraints mentioned above. For comparison, the Oklo natural reactor
constraints, which reflect the need for the $\mathrm{Sm}^{149}+n\rightarrow 
\mathrm{Sm}^{150}+\gamma $ neutron capture resonance at $97.3\,\mathrm{meV}$
to have been present $1.8-2\,\mathrm{Gyr}$ ($z=0.15$) ago, as first pointed
out by \cite{sh}, are currently \citep{fuj} $\Delta \alpha /\alpha =(-0.8\pm
1.0)\times 10^{-8}$ or $(8.8\pm 0.7)\times 10^{-8}$ (because of the
double-valued character of the neutron capture cross-section with reactor
temperature) and \citep{lam} $\Delta \alpha /\alpha >4.5\times 10^{-8} $ $%
(6\sigma )$ when the non-thermal neutron spectrum is taken into account.
However, there remain significant environmental uncertainties regarding the
reactor's early history and the deductions of bounds on constants. The
quoted Oklo constraints on $\alpha $ apply only when all other constants are
held to be fixed. If the quark masses to vary relative to the QCD scale, the
ability of Oklo to constrain variations in $\alpha $ is greatly reduced %
\citep{Flambaum07}.

Recently, Rosenband et al \cite{rosenband} measured the ratio of aluminium
and mercury single-ion optical clock frequencies, $f_{\mathrm{Al+}}/f_{%
\mathrm{Hg+}}$, over a period of about a year. From these measurements, the
linear rate of change in this ratio was found to be $(-5.3\pm 7.9)\times
10^{-17}\,\mathrm{yr}^{-1}$. These measurements provide the strongest limit
yet on any temporal drift in the value of $\alpha $: $\dot{\alpha}/\alpha
=(-1.6\pm 2.3)\times 10^{-17}\,\mathrm{yr}^{-1}$. This limit is strong
enough to conflict with theoretical explanations of the change in $\alpha $
reported by \cite{Webb:1999, Webb:2001} in terms of the slow variation of an
effectively massless scalar field, even allowing for the damping by
cosmological acceleration, unless there is a significant new physical effect
that slows the locally observed effects of changing $\alpha $ on
cosmological scales. Also, one expects inhomogeneous changes in $\alpha $ in
scenarios where variations in $\alpha $ are induced by a heavy scalar field
with a mass ($m_{\phi }$). One would expect variations in $\alpha $ on
cosmological scales to differ from those on scales below the field's Jeans
length, which is $O(1/m_{\phi })$ (for a detailed analysis of global-local
coupling of variations in constants see \cite%
{Clifton:2004,Mota:2004,Mota:2003,Shaw:2006a, Shaw:2006b, Shaw:2006c,
Olive:2007}).

It has also been noticed that if `constants' such as $\alpha$ or $\mu$ could
vary, then in addition to a slow temporal drift one would also expect to see
an annual modulation in their values. In many varying constant theories, the
Sun perturbs the values of the constants by a factor roughly proportional to
the Sun's Newtonian gravitational potential 
\citep{bmag, Barrow:2001,
Barrow:2002c} (the contribution from the Earth's gravitational potential is
about 14 times smaller than that of the Sun's at the Earth's surface). Hence
the `constants' depend on the distance from the Sun. Since the Earth's orbit
around the Sun has a small ellipticity, the distance, $r$, between the Earth
and Sun fluctuates annually, reaching a maximum at aphelion around the
beginning of July and a minimum at perihelion in early January. It was shown
that in many varying constant models, the values of the constants measured
here on Earth oscillate in a similar seasonal manner. Moreover, in many
cases, this seasonal fluctuation is predicted to dominate over any linear
temporal drift \citep{seasonal}.

In this paper we will study the \texttt{BSBM} model for the spatial and
temporal variations of the fine structure constant $\alpha $, with the aid
of full $N$-body simulations which explicitly and self-consistently solve
for the scalar field driving $\alpha$-evolution. We focus on the trend of
the scalar field (or equivalently, $\alpha$) inside the dark-matter halos,
and find that the profile of the scalar-field fluctuation is essentially
independent of the \texttt{BSBM} model parameter. This means that given the
density profile of an isolated halo and the background value of the scalar
field, we can accurately determine the scalar field perturbation in that
halo. We also derive an analytic expression for the scalar-field
perturbation using the Navarro-Frenk-White (NFW) halo profile, and show that
it agrees well with numerical results, at least for isolated halos. For
non-isolated halos this exact prediction differs from numerical result by a
(nearly) constant offset, depending on the environment of this halo.

A brief outline of the remaining of this paper is as follows: in \S ~\ref%
{sect:eqn} we list the minimal set of necessary equations to understand the
physics, and briefly describe our algorithm; \S ~\ref{sect:simu} displays
the main numerical results and we then discuss and conclude in \S ~\ref%
{sect:con}.

\section{Equations and Analysis}

\label{sect:eqn}

This section lists the equations which will be used in the $N$-body
simulations for the \texttt{BSBM} varying-$\alpha$ model \cite%
{Barrow:2002b,Barrow:2002,Barrow:2001,Sandvik:2001,Mota:2003,Mota:2004}.

\subsection{The Basic Equations}

\label{subsect:basiceqn}

The Lagrangian density for the \texttt{BSBM} model could be written as 
\begin{eqnarray}  \label{eq:Lagrangian}
\mathcal{L} &=& \frac{1}{2}\left[ \frac{R}{\kappa }-\nabla ^{a}\varphi
\nabla _{a}\varphi \right]-\mathcal{L}_{m}- e^{-2\sqrt{\kappa}\varphi}%
\mathcal{L}_{\mathrm{EM}}-\mathcal{L}_{r}\ ,\ 
\end{eqnarray}
where $R$ is the Ricci scalar, $\kappa =8\pi G$ with $G$ being the
gravitational constant, $\varphi $ is the scalar field; $\mathcal{L}_{m}, 
\mathcal{L}_{\mathrm{EM}}, \mathcal{L}_{r}$ represent respectively the
Lagrangian densities for dust, electromagnetic field (including photons) and
other radiation (such as neutrinos). The coupling function between the
scalar field and the electromagnetic field in the \texttt{BSBM} model is $%
e^{-2\sqrt{\kappa}\varphi}$ where $\sqrt{\kappa}$ is added so that $\sqrt{%
\kappa}\varphi\equiv\psi$ is dimensionless. In the simplest version of the
model there is no potential for the scalar field.

The dust Lagrangian for a point particle with mass $m_{0}$ is 
\begin{eqnarray}  \label{eq:DMLagrangian}
\mathcal{L}_{m}(\mathbf{y}) &=& -\frac{m_{0}}{\sqrt{-g}}\delta (\mathbf{y}-%
\mathbf{x}_{0})\sqrt{g_{ab}\dot{x}_{0}^{a}\dot{x}_{0}^{b}},
\end{eqnarray}
where $\mathbf{y}$ is the general coordinate and $\mathbf{x}_{0}$ is the
coordinate of the centre of the particle. From this equation we derive the
corresponding energy-momentum tensor: 
\begin{eqnarray}  \label{eq:DMEMT_particle}
T_{m}^{ab} &=& \frac{m_{0}}{\sqrt{-g}}\delta (\mathbf{y}-\mathbf{x}_{0})\dot{%
x}_{0}^{a}\dot{x}_{0}^{b}.
\end{eqnarray}
Also, because $g_{ab}\dot{x}_{0}^{a}\dot{x}_{0}^{b}\equiv
g_{ab}u^{a}u^{b}=1, $ in which $u^{a}$ is the four-velocity of the
dark-matter particle centred at $x_{0}$, the Lagrangian can be rewritten as 
\begin{eqnarray}  \label{eq:DMLagrangian2}
\mathcal{L}_{m}(\mathbf{y}) &=& -\frac{m_{0}}{\sqrt{-g}}\delta(\mathbf{y}-%
\mathbf{x}_{0}).
\end{eqnarray}
This result will be used below.

Eq.~(\ref{eq:DMEMT_particle}) is just the energy-momentum tensor for a
single matter particle. For a fluid consisting of many particles, the
energy-momentum tensor will be 
\begin{eqnarray}  \label{eq:DMEMT_fluid}
T_{m}^{ab} &=& \frac{1}{\mathcal{V}}\int_{\mathcal{V}}d^{4}y\sqrt{-g}\frac{%
m_{0}}{\sqrt{-g}}\delta (y-x_{0})\dot{x}_{0}^{a}\dot{x}_{0}^{b}  \notag \\
&=&\rho_{\mathrm{CDM}}u^{a}u^{b},
\end{eqnarray}%
where $\mathcal{V}$ denotes a volume which is microscopically large but
macroscopically small, and we have extended the 3-dimensional $\delta $
function to a 4-dimensional one by adding a time component. Here, $u^{a}$ is
the averaged four-velocity of the matter fluid.

Using 
\begin{eqnarray}
T^{ab} &=& -\frac{2}{\sqrt{-g}}\frac{\delta\left(\sqrt{-g}\mathcal{L}\right)%
}{\delta g_{ab}},
\end{eqnarray}
it is straightforward to show that the energy-momentum tensor for the scalar
field is 
\begin{eqnarray}  \label{eq:phiEMT}
T^{\varphi ab} &=& \nabla^{a}\varphi\nabla^{b}\varphi -\frac{1}{2}%
g^{ab}\nabla_{c}\varphi\nabla^{c}\varphi .
\end{eqnarray}
Therefore the total energy-momentum tensor is 
\begin{eqnarray}  \label{eq:EMT_tot}
T_{ab} &=& \nabla_{a}\varphi\nabla_{b}\varphi- \frac{1}{2}%
g_{ab}\nabla_{c}\varphi\nabla^{c}\varphi  \notag \\
&& +T^{m}_{ab}+T^{r}_{ab} +e^{-2\sqrt{\kappa}\varphi}T_{ab}^{\mathrm{EM}}
\end{eqnarray}%
where $T_{ab}^{m}=\rho _{m}u_{a}u_{b}$, $T_{ab}^{\mathrm{r}}$ is the
energy-momentum tensor for radiation fields except photons, and $T^{\mathrm{%
EM}}_{ab}$ for photons. The Einstein equations are 
\begin{eqnarray}  \label{eq:EinsteinEq}
G_{ab} &=& \kappa T_{ab}
\end{eqnarray}
in which $G_{ab}=R_{ab}-\frac{1}{2}g_{ab}R$ is the Einstein tensor. Note
that due to the extra coupling between the scalar field, $\varphi$, and the
electromagnetic field, the energy-momentum tensors for either will no longer
be separately conserved, but instead we have 
\begin{eqnarray}  \label{eq:DM_energy_conservation}
\nabla _{b}T_{\mathrm{EM}}^{ab} &=& 2\sqrt{\kappa}\left( g^{ab}\mathcal{L}_{%
\mathrm{EM}}+T_{\mathrm{EM}}^{ab}\right) \nabla _{b}\varphi.
\end{eqnarray}
However, the total energy-momentum tensor is certainly conserved.

Meanwhile, the scalar field equation of motion is 
\begin{eqnarray}  \label{eq:phiEOM}
\square \varphi &=& -2\sqrt{\kappa}e^{-2\sqrt{\kappa}\varphi}\mathcal{L}_{%
\mathrm{EM}},
\end{eqnarray}
where $\square\equiv\nabla^{a}\nabla_{a}$. This equation governs the time
evolution and spatial configuration of the scalar field.

Eqs.~(\ref{eq:EMT_tot}, \ref{eq:EinsteinEq}, \ref{eq:DM_energy_conservation}%
, \ref{eq:phiEOM}) summarize all the physics needed for the following
analysis. However, when making use of them we should also specify the form
of the electromagnetic matter. For example, if it is a pure electromagnetic
field (photons), then we have $\mathcal{L}_{\mathrm{EM}}=\frac{1}{2}%
\left(E^2-B^2\right)=0$ in which $E, B$ stand for the electric and magnetic
fields. Thus from the time component of Eq.~(\ref{eq:DM_energy_conservation}%
) we obtain the (background) evolution equation for photon density as 
\begin{eqnarray}
\dot{\rho}_r + 4H\rho_r &=& 2\dot{\psi}\rho_{r}
\end{eqnarray}
where remember that $\psi=\sqrt{\kappa}\varphi$.

It then seems that, by the same reason, the right hand side of Eq.~(\ref%
{eq:phiEOM}) also vanishes, leaving the scalar field unsourced. This may not
be true however, as non-relativistic matter could also contribute to $%
\mathcal{L}_{\mathrm{EM}}$ and thus $T_{ab}^{\mathrm{EM}}$. For example, in
baryonic matter $\mathcal{L}_{\mathrm{EM}}\approx E^{2}/2$, and for neutrons
and protons this electromagnetic contribution to the total mass can be of
order $10^{-4}$; in superconducting cosmic strings $\mathcal{L}_{\mathrm{EM}%
}\approx -B^{2}/2$ where $\rho _{\mathrm{EM}}\approx B^{2}/2$ so that $|%
\mathcal{L}_{\mathrm{EM}}/\rho _{\mathrm{EM}}|\sim 1$. In the \texttt{BSBM}
model, in order to simplify the situation, it is assumed that $\mathcal{L}_{%
\mathrm{EM}}/\rho _{m}=\zeta $ where $\zeta $ is a constant with a modulus
between $0$ and $\approx 1$, either positive or negative, and $\rho _{m}$ is
the density for non-relativistic matter.

Thus the scalar field equation gets sourced by a term proportional to $\zeta$%
: 
\begin{eqnarray}  \label{eq:phiEOM2}
\square \varphi &=& -2\sqrt{\kappa}\zeta e^{-2\sqrt{\kappa}\varphi}\rho_{m}.
\end{eqnarray}

Since the part of $\mathcal{L}_{\mathrm{EM}}$ which affects the scalar field
is a constituent of the non-relativistic matter and is presumably moving
with the matter particles, we could combine Eq.~(\ref%
{eq:DM_energy_conservation}) and the conservation equation for the dust
matter (no including electromagnetic contribution) to write a new
conservation equation for the particle: 
\begin{eqnarray}
\nabla_{b}T_{m+\mathrm{EM}}^{ab} &=& 2\sqrt{\kappa}\left( g^{ab}\mathcal{L}_{%
\mathrm{EM}}+T_{\mathrm{EM}}^{ab}\right) \nabla _{b}\varphi.
\end{eqnarray}
Although we have assumed above that $\mathcal{L}_{\mathrm{EM}}=\zeta\rho_m$,
we still lack a knowledge about $T_{\mathrm{EM}}^{ab}$, whose relation to $%
\mathcal{L}_{\mathrm{EM}}$ could be complicated. Here for simplicity we
assume that $T^{ab}_{\mathrm{EM}}=-\zeta\rho_{m}u^{a}u^b$. Then it is easy
to find that the time component of this equation reads 
\begin{eqnarray}
\dot{\rho}_{m+\mathrm{EM}}+3H\rho_{m+\mathrm{EM}} &=& 0
\end{eqnarray}
while the $i$-th spatial component of it gives the following (modified)
geodesic equation 
\begin{eqnarray}  \label{eq:geodesic}
\ddot{x}^{i}_{0} + \Gamma^{i}_{ab}\dot{x}^{a}_{0}\dot{x}^{b}_{0} &=& 2\zeta%
\sqrt{\kappa} \left(g^{ib}-u^{i}u^{b}\right)\nabla_{b}\varphi.
\end{eqnarray}
where $x_0$ is the coordinate of the centre of a particle, and the right
hand side represents a fifth force on the particle \cite{Li:2009, Li:2010}.
The assumption $T^{ab}_{\mathrm{EM}}=-\zeta\rho_{m}u^{a}u^b$ might seem
unappealing, but as we shall see below, because $|\zeta|\ll1$, the fifth
force is much weaker than gravity and thus has negligible effects in the
clustering of matter in any case; ultimately it is only the \texttt{BSBM}
assumption $\mathcal{L}_{\mathrm{EM}}=\zeta\rho_m$ that is important in
theoretical predictions of the spatial and temporal variations of $\alpha$,
given that $\alpha=e^{2\psi}\frac{e^{2}_0}{c\hbar}$.

\subsection{Analytical Approximation}

\label{subsect:approx}

The scalar field equation of motion, which controls the dynamics of the
scalar field $\varphi$, is generally complicated, because it depends
nonlinearly on $\varphi$, which both evolves in time and fluctuates in
space. Fortunately, for the majority of applications the scalar field
potential and coupling function are not nonlinear enough to give the scalar
field a very heavy mass so as to make it fluctuate strongly. The nice thing
about this is that in certain places of the equations one may then forget
the scalar field perturbation and simplify these equations accordingly. In 
\cite{Li:2010b}, for example, it is shown that such simplification is very
good an approximation (see however \cite{Li:2009, Li:2010} for an opposite
extreme for which the scalar field potential is very nonlinear so that such
simplification does not work).

In the \texttt{BSBM} model, there is no scalar field potential and the
coupling function is close to linear if $\sqrt{\kappa}|\varphi|\ll1$ (which
is the case for our interested parameter space \citep{Barrow:2002}). We
therefore expect the fluctuation of $\varphi$ to be very weak and assume
that $\sqrt{\kappa}|\delta\varphi|\ll\sqrt{\kappa}|\varphi|\ll1$ (which we
shall confirm below using numerical simulations). Under such an assumption
the Poisson equation could be written as 
\begin{eqnarray}  \label{eq:Poisson}
\nabla_{\mathbf{x}}^{2}\Phi &=& 4\pi Ga^{3}\rho_m\left[1+\zeta e^{-2\sqrt{%
\kappa}\left(\bar{\varphi}+\delta\varphi\right)}\right]  \notag \\
&&- 4\pi Ga^{3}\bar{\rho}_m\left[1+\zeta e^{-2\sqrt{\kappa}\bar{\varphi}}%
\right]  \notag \\
&\approx& 4\pi Ga^3\left(\rho_{m}-\bar{\rho}_m\right)
\end{eqnarray}
where $\bar{\varphi}$ the background value of $\varphi$ and $\delta\varphi$
its perturbation; $\rho_m, \bar{\rho}_m$ are respectively the local and
background matter density; $\Phi$ is the gravitational potential and $a$ the
cosmic scale factor; $\nabla_{\mathbf{x}}$ is the derivative with respect to
the comoving coordinate $\mathbf{x}$. To obtain Eq.~(\ref{eq:Poisson}) we
have used the fact that in the \texttt{BSBM} model $|\zeta e^{-2\bar{\varphi}%
}|\ll1$. Note that Eq.~(\ref{eq:Poisson}) clearly indicates that the
gravitational potential is essentially not influenced by the scalar field $%
\varphi$.

For the scalar field equation of motion Eq.~(\ref{eq:phiEOM2}), because the
background part (which has no spatial dependence) can be solved easily, so
we subtract that from the full equation to obtain an equation of motion for $%
\delta\varphi$ only (remember that $\varphi=\bar{\varphi}+\delta\varphi$).
Furthermore, we drop the time derivative terms of $\delta\varphi$ as they
are small compared with the spatial gradients (\emph{i.e.}, work in the
quasi-static limit). The final equation for $\delta\varphi$ then becomes 
\begin{eqnarray}  \label{eq:phiEOM3}
\nabla^{2}_{\mathbf{x}}\left(a\sqrt{\kappa}\delta\varphi\right) &=&
2\zeta\kappa\left[\rho_me^{-2\sqrt{\kappa}\left(\bar{\varphi}+
\delta\varphi\right)}-\bar{\rho}_me^{-2\sqrt{\kappa}\bar{\varphi}}\right]a^3
\notag \\
&\approx& 16\pi Ga^3\zeta\left(\rho_m-\bar{\rho}_m\right)\ \ 
\end{eqnarray}
where we have used $\kappa=8\pi G$ and $\sqrt{\kappa}|\delta\varphi|\ll\sqrt{%
\kappa}|\varphi|\ll1$.

Comparing Eqs.~(\ref{eq:Poisson}, \ref{eq:phiEOM3}), it is evident that the
source terms are the same up to a constant coefficient $4\zeta$.
Consequently, we shall have 
\begin{eqnarray}  \label{eq:ratio}
a\sqrt{\kappa}\delta\varphi(\mathbf{x}) &\approx& 4\zeta\Phi(\mathbf{x}).
\end{eqnarray}
Note that this equation, together with the geodesic equation Eq.~(\ref%
{eq:geodesic}), implies that the magnitude of the fifth force (force due to
exchange of scalar field quanta between particles) $|\mathbf{f}|$ satisfies 
\begin{eqnarray}
|\mathbf{f}|\ \sim\ \zeta|\vec{\nabla}(a\sqrt{\kappa}\delta\varphi)|\ \sim\
4\zeta^2|\vec{\nabla}\Phi|.
\end{eqnarray}
Therefore the ratio between the magnitudes of the fifth force and gravity is
of order $\zeta^2\lesssim10^{-12}-10^{-8}\ll1$. This implies that the fifth
force cannot influence the structure formation significantly.

\section{Simulations and Results}

\label{sect:simu}

\subsection{$\protect\varphi$ Perturbation vs.~Gravitational Potential}

\begin{figure}[tbp]
\centering \includegraphics[scale=0.49] {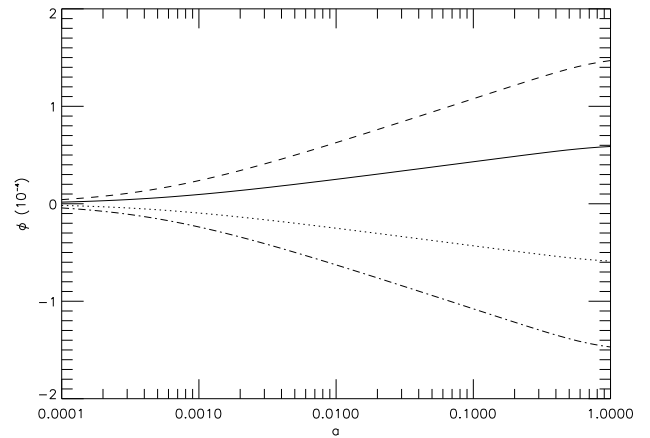}
\caption{The time evolution of $\bar{\protect\varphi}$ for the four models
considered in this work, with $\protect\zeta =-2\times 10^{-6}$ (solid
curve), $2\times 10^{-6}$ (dotted curve), $-5\times 10^{-6}$ (dashed curve)
and $5\times 10^{-6}$ (dash-dotted curve) respectively. The horizontal axis
is the cosmic scale factor $a(t)$ and the vertical axis plots $\protect\sqrt{%
\protect\kappa }\bar{\protect\varphi}$ in unit of $10^{-4}$.}
\label{fig:background}
\end{figure}

\begin{figure*}[tbp]
\centering \includegraphics[scale=1.7] {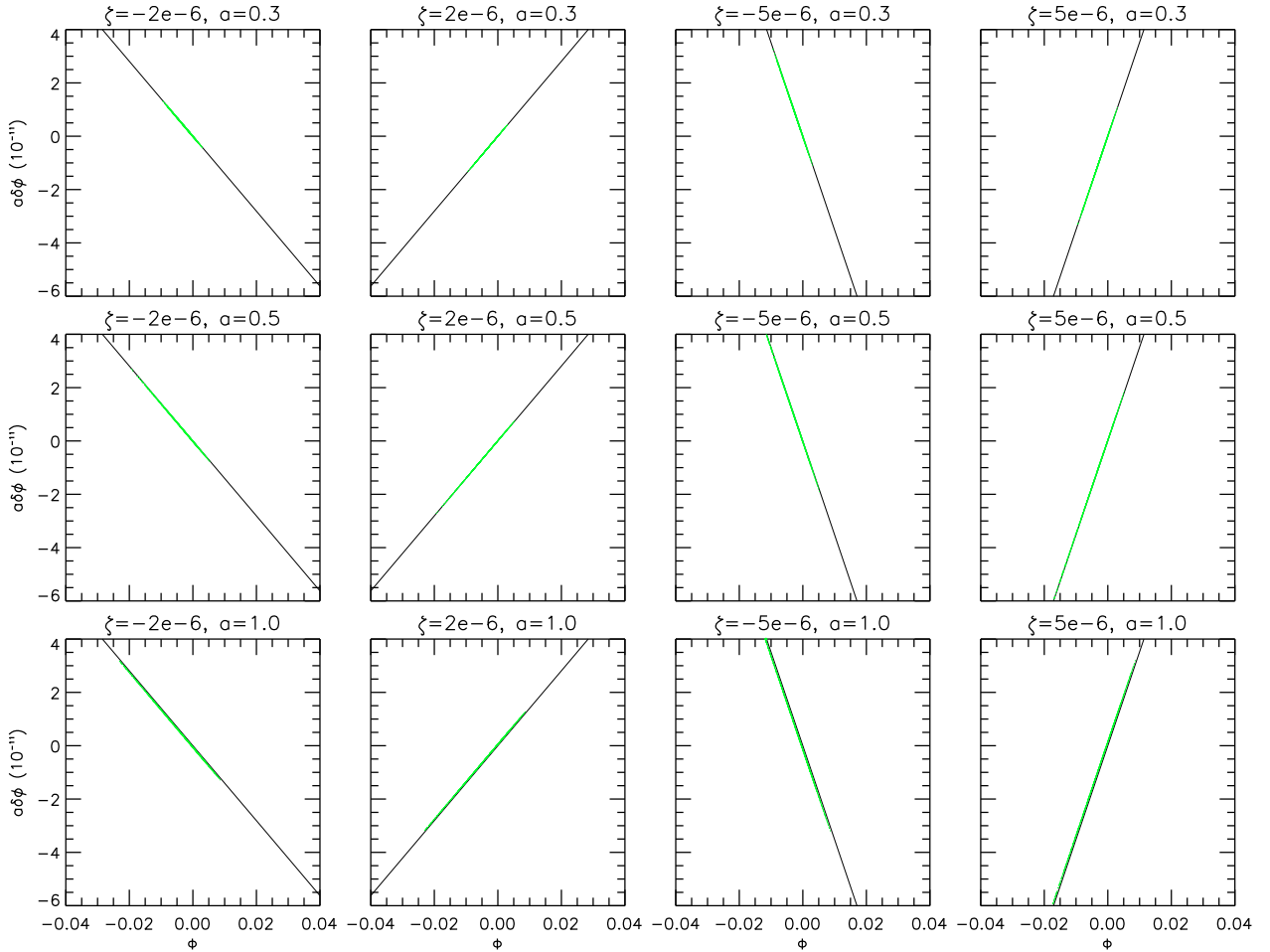}
\caption{(Color Online) Scatter plot of the scalar field perturbation, $a%
\protect\sqrt{\protect\kappa }\protect\delta \protect\varphi ,$ (vertical
axis, in unit of $10^{-11}$) vesus gravitational potential $\Phi $
(horizontal axis) in code units. The black solid line is the analytical
approximation $a\protect\delta \protect\varphi \propto \Phi $; each green
dot represents the corresponding result measured in a cell from a particular
slice that is randomly selected from the simulation box. The columns are for
four models with $\protect\zeta =\pm 2,\pm 5\times 10^{-6}$ as shown above
each panel, and the rows are for three different output times $a=0.3,0.5,1.0$%
, also shown above each panel. Note that the slopes for the solid lines
differ because of the different values of $\protect\zeta $.}
\label{fig:scal_pot}
\end{figure*}

\begin{figure*}[tbp]
\centering \includegraphics[scale=1.9] {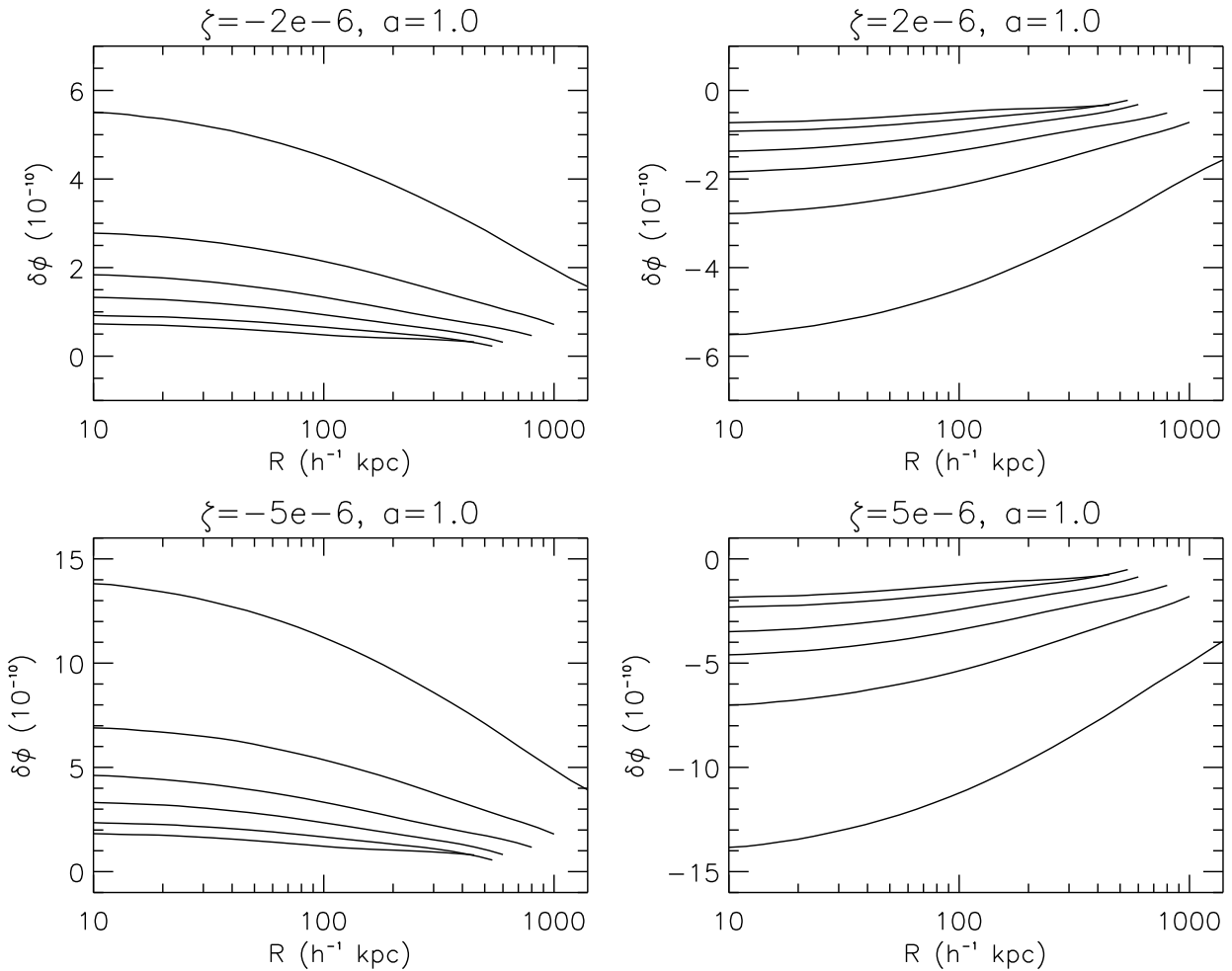}
\caption{The profile of $a\protect\sqrt{\protect\kappa }\protect\delta 
\protect\varphi $ inside the halos for the four models with $\protect\zeta %
=\pm 2,\pm 5\times 10^{-6}$ as indicated at the top of each panel. We have
considered the 80 most massive halos from the simulation box, and divided
them into six bins with $M/\left( 10^{14}M_{\bigodot }\right) \in \lbrack
2.0,\infty )$, $[1.0,2.0],[0.6,1.0],[0.3,0.6],[0.2,0.3]$ and $[0.1,0.2]$
respectively. Each curve represents the averaged profile of $a\protect\sqrt{%
\protect\kappa }\protect\delta \protect\varphi $ for one bin (the more
massive bins always correspond to larger, either positive or negative --
depending on the sign of $\protect\zeta $ -- deviations from zero). All
results are for output time $a=1.0$ (today). The horizontal axis is the
distance $R$ to the halo centre, in units of $h^{-1}~$kpc and the vertical
axis is $a\protect\sqrt{\protect\kappa }\protect\delta \protect\varphi $ in
units of $10^{-10}$.}
\label{fig:profile_scal}
\end{figure*}

\begin{figure*}[tbp]
\centering \includegraphics[scale=1.9] {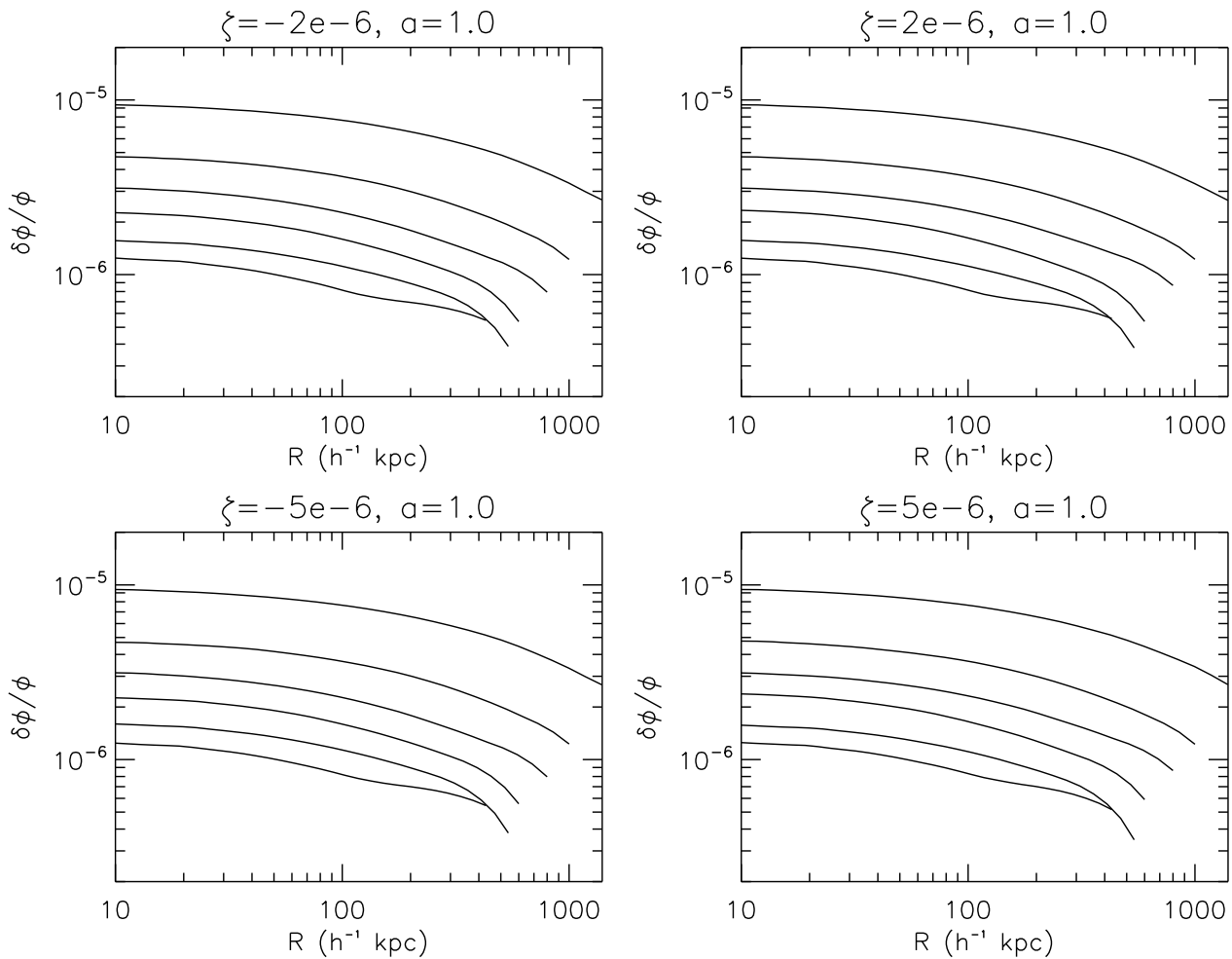}
\caption{As Fig.~\protect\ref{fig:profile_scal}, except that this figure
shows the profiles of $\protect\delta \protect\varphi /\protect\varphi $
instead of $a\protect\sqrt{\protect\kappa }\protect\delta \protect\varphi $.}
\label{fig:profile_scal2}
\end{figure*}

\begin{figure*}[tbp]
\centering \includegraphics[scale=0.9] {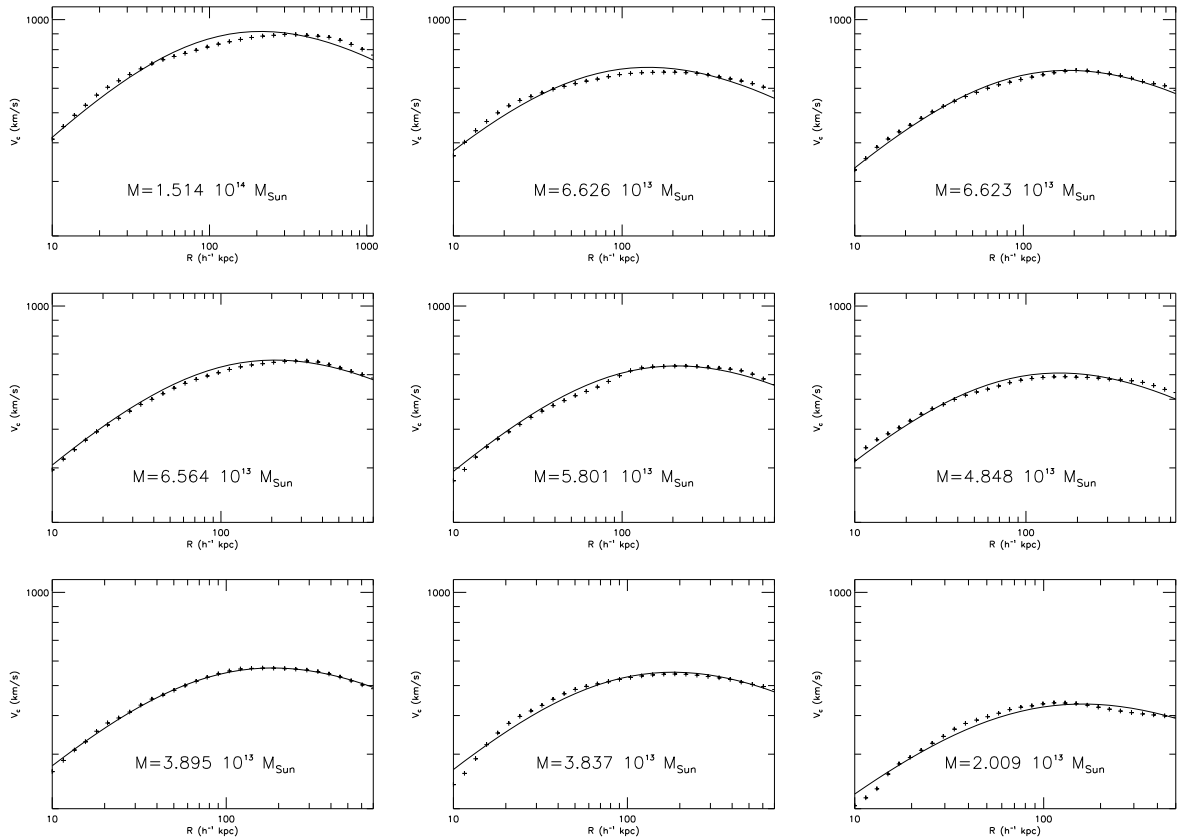}
\caption{The fitting curves for the circular velocity, $V_{c},$ in the halo
using the parameterization Eq.~(\protect\ref{eq:V_c}). We show the results
for nine halos selected from the 80 most massive ones in the simulation box,
and their masses are given near the bottom of the each panel (the mass is
decreasing from the upper-left corner to the lower-right corner). Solid
curves are the fitting formulae while crosses are the $N$-body simulation
results. All results are at the output time $a=1.0$; the horizontal axis is
the distance from the halo centre in unit of $h^{-1}~$kpc and vertical axis
denotes $V_{c}$, in unit of km/s. Note that only the model with $\protect%
\zeta =-2\times 10^{-6}$ is displayed for clarity but other models give
similar results.}
\label{fig:profile_vc}
\end{figure*}

\begin{figure*}[tbp]
\centering \includegraphics[scale=0.9] {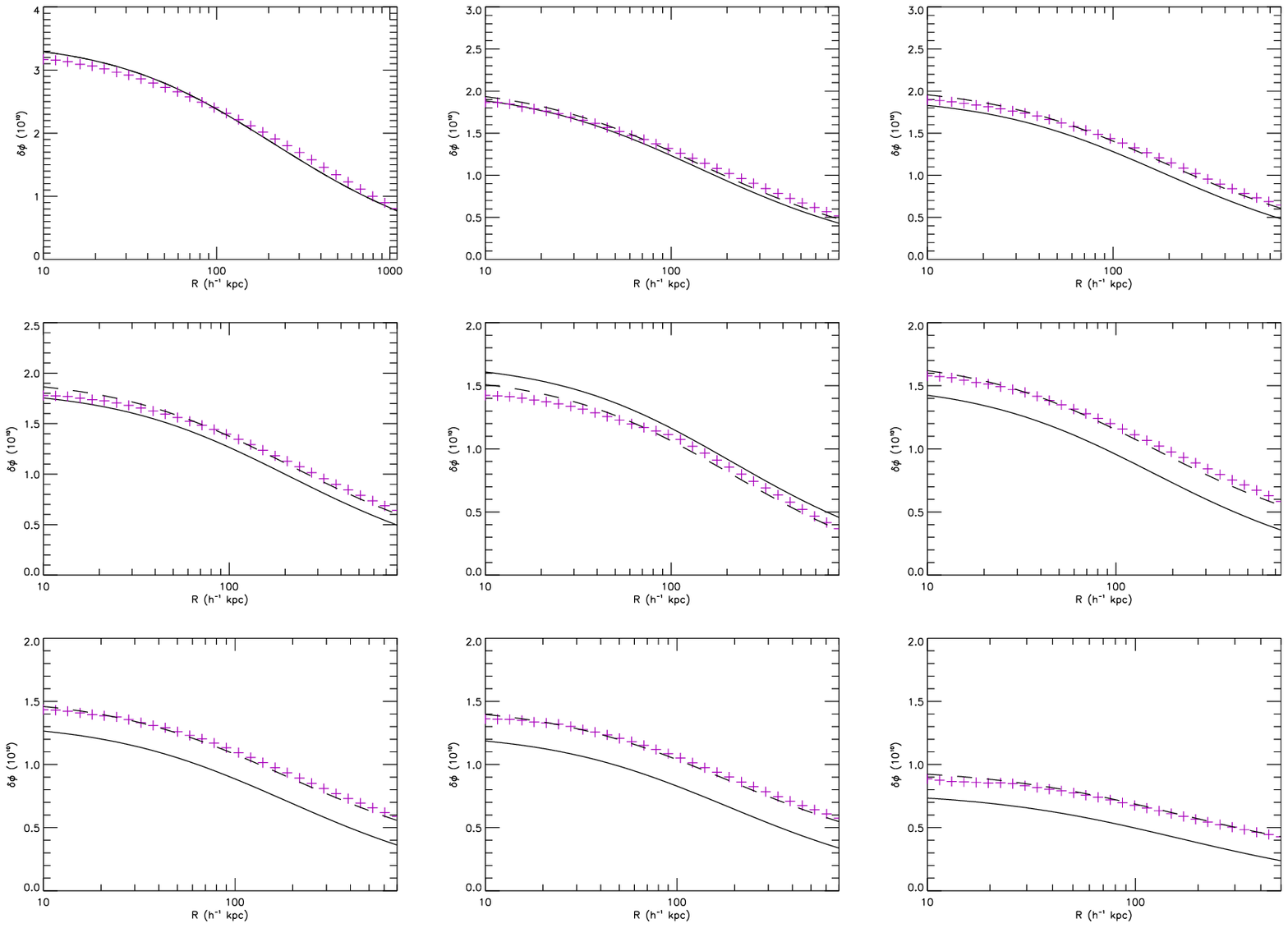}
\caption{(Color Online) The nalytic approximation compared with the
numerical simulation results for the profile of $a\protect\sqrt{\protect%
\kappa }\protect\delta \protect\varphi $ in the same nine halos as in Fig.~%
\protect\ref{fig:profile_vc}. Purple crosses are the numerical results, the
solid curve represents the analytical approximation Eq.~(\protect\ref%
{eq:delta_varphi}) with $\Phi _{\ast }=\Phi _{\infty }=0$, and the dashed
curve denotes Eq.~(\protect\ref{eq:delta_varphi}) with some tuned value of $%
\Phi _{\ast }$. The parameters $\protect\beta $ and $R_{s}$ are the best-fit
values in Fig.~\protect\ref{fig:profile_vc}. All results are at the output
time when $a=1.0$; the horizontal axis is the distance from the halo centre,
in units of $h^{-1}~$kpc and the vertical axis denotes $a\protect\sqrt{%
\protect\kappa }\protect\delta \protect\varphi $, in units of $10^{-10}$.
Note that only the model with $\protect\zeta =-2\times 10^{-6}$ is displayed
for clarity but other models give similar results.}
\label{fig:profile_scalfit}
\end{figure*}

To study in details the behaviour of the scalar field and hence the fine
structure constant $\alpha$ in the \texttt{BSBM} model, we have performed $N$%
-body simulations for four different models, with $\zeta=\pm2\times10^{-6}$
and $\pm5\times10^{-6}$ respectively. The physical parameters we adopt in
all simulations are as follows: the present-day dark-energy fractional
energy density $\Omega_{\mathrm{DE}}=0.743$ and $\Omega_{m}=\Omega_{\mathrm{%
CDM}}+\Omega_{\mathrm{B}}=0.257$, $H_{0}=71.9$~km/s/Mpc, $n_{s}=0.963$, $%
\sigma_{8}=0.761$. These are in accordance with the concordance cosmological
model preferred by current data sets. Our simulation box has a size of $%
64h^{-1}$~Mpc, in which $h=H_{0}/(100~\mathrm{km/s/Mpc})$. In all those
simulations, the mass resolution is $1.114\times 10^{9}h^{-1}~M_{\bigodot }$%
; the particle number is $256^{3}$; the domain grid (\emph{i.e.}, the
coarsest grid which covers the whole simulation box) has $128^{3}$
equal-sized cubic cells and the finest refined grids have 16384 cells on
each side, corresponding to a force resolution of $\sim12h^{-1}~$kpc.
Detailed description about the $N$-body simulation technique and code for
the (coupled) scalar field models could be found in \cite{Li:2009, Li:2010}
and will not be presented here.

Because the \texttt{BSBM} model (with the parameter $\zeta$ constrained by
data) involves a very weak coupling between matter and the scalar field $%
\varphi$, the presence of the latter and its coupling have negligible
influences on the background ($\Lambda$CDM) cosmology, although the opposite
is not true since there is no scalar field potential and thus the dynamics
of $\varphi$ is controlled entirely by the coupling. In our simulations, we
compute the \emph{full} background cosmology and evolution of $\varphi$ on a
predefined time grid using \texttt{MAPLE}, and interpolate to obtain the
corresponding quantities which are needed in $N$-body simulations. Details
of this procedure can be found in the Appendix C of \cite{Li:2010b}, and
Fig.~\ref{fig:background} shows the background evolution of $\sqrt{\kappa}%
\varphi$ for the 4 models considered here. Obviously the condition $\sqrt{%
\kappa}\bar{\varphi}\ll1$ is satisfied, justifying the approximation we used
above to derive Eq.~(\ref{eq:ratio}).

One nice thing about the \texttt{BSBM} model is its simplicity, and it turns
out that the background evolution of $\varphi $ (and thus $\alpha $) in
different cosmic epochs can be well described by some analytical formulae 
\cite{Barrow:2002}. Therefore in the present study we shall mainly focus on
the spatial variation of $\varphi $ and $\alpha $ (especially in virialized
halos).

As mentioned above, because there is no potential for the scalar field $%
\varphi$ and because $\sqrt{\kappa}\varphi\ll1$ for our choices of $\zeta$,
so the scalar field equation of motion (in the quasi-static limit) for $%
\delta\varphi$ and the Poisson equation share the same source up to a
constant coefficient, and therefore we expect $a\delta\varphi\propto\Phi$
across the whole space. This is what we have found in \cite{Li:2010b} for a
different coupled scalar field model where the scalar field potential is
negligible. Indeed, this could serve as a test of the scalar field solver in
our $N$-body simulation code.

To check that our code does recover the analytical approximation, we have
plotted in Fig.~\ref{fig:scal_pot} the comparison of the scalar field
perturbation $a\sqrt{\kappa}\delta\varphi$ and gravitational potential $\Phi$
from a slice of the simulation box. As indicated by this figure, the
agreement between the numerical results (green dots) and analytical
approximation (black solid line) is remarkably good, implying that the
numerical code works well. Therefore to a high precision we can assume that $%
a\sqrt{\kappa}\delta\varphi\propto\Phi$ everywhere, a fact which shall be
used below to obtain an analytical expression of $\delta\varphi$ in dark
matter halos.

\subsection{Spatial variation of $\protect\varphi$ in Halos}

In the standard picture, the galaxies where observers live generally locate
inside the dark matter halos, which to the simplest approximation are just
spherical clusters of matter with a universal NFW \citep{NFW} density
profile.

We are certainly interested in the (possible) variation of $\alpha$ inside
the halo we reside in. For example, there has been a great deal of
analytical work on how significantly the local value of $\alpha$ could
deviate from its cosmological counterpart 
\citep{Mota:2003,Mota:2004,Jacobson:1999, Wetterich:2003, Shaw:2006a, Shaw:2006b,
Shaw:2006c}. Also, if the spatial variation of $\alpha$ is strong enough,
then it might have impact on our observation of the spectra for the stars
from our Galaxy and other galaxies.

From our simulation output, it is easy to identify the dark matter halos %
\citep{mhf, Li:2010b}. Now what we want to do here is to measure the
quantity $\sqrt{\kappa}\delta\varphi$ as a function of distance $R$ to the
halo centres, assuming that the halos are exactly spherical. For this we
have recorded the value for $\sqrt{\kappa}\delta\varphi$ at the position of
each particle, and then presumably we could divide each halo into a number
of spherical shells, determine the radius of each shell and compute the
average value of $\sqrt{\kappa}\delta\varphi$ in the shells. However, there
is some subtlety in the computation of the averaged $\sqrt{\kappa}%
\delta\varphi$.

The problem is that, since we have only recorded the information for $\sqrt{%
\kappa}\delta\varphi$ at the positions of the particles, we do not have a
fair sampling points. Because $\sqrt{\kappa}\delta\varphi$ tends to be
different in regions of different particle number densities, the high
density regions will be over-sampled and low density regions under-sampled,
resulting in a bias as we are trying to determine the \emph{spatially}
averaged, rather than the \emph{particle} averaged value, of $\sqrt{\kappa}%
\delta\varphi$. To test how big the bias could be, we use an approximation
as follows: firstly we divide each spherical shell into $N$ equal-sized
volumes which are small enough so that the particle number density do not
change much inside each of them, then we compute the \emph{particle} average
of $\sqrt{\kappa}\delta\varphi$ in each of these volumes, call it $\langle%
\sqrt{\kappa}\delta\varphi\rangle_{i}$ for $i=1\cdots N$, then the spatial
(volume) average of $\sqrt{\kappa}\delta\varphi$ is given by 
\begin{eqnarray}  \label{eq:vol_aver}
\langle\sqrt{\kappa}\delta\varphi\rangle_{\mathrm{Vol}} &\approx& \frac{1}{N}%
\sum_{i=1}^{N}\langle\sqrt{\kappa}\delta\varphi\rangle_{i}.
\end{eqnarray}
More precise treatments of the volume average could be obtained by using
space tessellations, such as Delaunay triangulation, but this is too
technical and thus beyond the scope of this work. Anyway, using our
approximation Eq.~(\ref{eq:vol_aver}), we find that the bias caused by using
particle-number average is at most $1\sim2\%$, which is not unacceptable
considering that the sphericity of halos is already an approximation.

Fig.~\ref{fig:profile_scal} shows the profile of $\sqrt{\kappa}\delta\varphi$
inside the dark matter halos. Instead of plotting this halo by halo, we have
selected the 80 most massive halos from our simulation box, divide them into
six bins with $M/\left(10^{14}M_{\bigodot}\right)\in[2.0,\infty)$, $%
[1.0,2.0] $, $[0.6,1.0]$, $[0.3,0.6]$, $[0.2,0.3]$ and $[0.1,0.2]$. Then we
compute the averaged profile of $\sqrt{\kappa}\delta\varphi$ in each of the
six bins (the six curves in Fig.~\ref{fig:profile_scal}). As expected, the
larger the halo is, the deeper the gravitational potential is and, because $a%
\sqrt{\kappa}\delta\varphi\approx\Phi$, the larger $|\sqrt{\kappa}\varphi|$
is. Meanwhile, going from the halo centre outwards one finds that $|\sqrt{%
\kappa}\varphi|$ gradually decreases (towards $0$), which makes sense
because $\sqrt{\kappa}\delta\varphi\rightarrow0$ as $R\rightarrow\infty$.
Finally, we also notice that the value of $\sqrt{\kappa}\delta\varphi$
depends sensitively on $\zeta$ and $|\sqrt{\kappa}\delta\varphi|$ increases
with $|\zeta|$.

Fig.~\ref{fig:profile_scal} also shows that for our choices of $\zeta $ the
value of $|\sqrt{\kappa }\delta \varphi |$ is typically of order $%
10^{-10}\sim 10^{-9}$, which is quite small but gives us no idea how large
the fluctuation in $\varphi $ is. For the latter we have instead plotted the
profile of the quantity $\delta \varphi /\varphi $ inside the 80 halos
distributed in six bins. Interestingly, unlike $\delta \varphi $, the
quantity $\delta \varphi /\varphi $ does not depend on the sign of $\zeta $,
and indeed it almost does not depend on the magnitude of $\zeta $ either!
This, together with the facts that (a) $a\sqrt{\kappa }\delta \varphi
(R,a)\propto \Phi (R,a)$ and (b) the presence of the scalar field and its
coupling to matter have negligible effect in the structure formation (so
that the halo density profile remains NFW), indicates that the fluctuation
of $\varphi $ in the \texttt{BSBM} model only depends on the \emph{non}-%
\texttt{BSBM} physical parameters, and once we have solved the background
value of $\varphi $ we might get some pretty good idea about the $\sqrt{%
\kappa }\delta \varphi $ profile without solving it explicitly (see below
for further details). Back to the size of $\delta \varphi /\varphi $, from
Fig.~\ref{fig:profile_scal2} we see clearly that it is of order $10^{-6}\sim
10^{-5}$, which is too tiny to produce any observable effects in the spatial
variation of $\alpha $.

Now, given that the NFW profile for dark matter halos is (expected to be)
preserved in the varying-$\alpha$ simulations, we wonder whether it is
possible to derive some analytical (approximate) formula for the profile of $%
\sqrt{\kappa}\varphi$. For this let's recall that the NFW profile is
expressed as 
\begin{eqnarray}  \label{eq:NFW}
\frac{\rho(r)}{\rho_{c}} &=& \frac{\beta}{\frac{r}{R_s}\left(1+\frac{r}{R_s}%
\right)^2}
\end{eqnarray}
where $\rho_c$ is the critical density for matter, $\beta$ is a
dimensionless fitting parameter and $R_s$ a second fitting parameter with
length dimension. $\beta$ and $R_s$ are generally different for different
halos and should be fitted for individual halos.

We have checked the halos in our analysis and found that the majority of
them are indeed very well described by Eq.~(\ref{eq:NFW}), confirming our
earlier argument that the coupled scalar field effect is too tiny to change
the structure formation \footnotemark[1]. However, we shall not use the
fitting to Eq.~(\ref{eq:NFW}) in this work, mainly for two reasons: first,
the dark matter density profile is in general difficult to measure directly
or precisely, while in contrast the circular velocities $V_{c}$ of the stars
rotating about the halo centre are easier to measure; second, $V_{c}$ as a
function of radius $R$ is more closely related to $\sqrt{\kappa }\delta
\varphi (R)$, which will become clear later. As a result, we shall use
fittings to $V_{c}$ from here on.

\footnotetext[1]{%
Indeed the NFW profile is quite robust, and even the scalar field does
influence the structure formation significantly it is often still preserved.
See examples for the coupled quintessence \citep{Baldi:2010}, ReBEL %
\citep{Keselman:2010} and extended quintessence \citep{LMB:2010} models.}

Assuming Eq.~(\ref{eq:NFW}) as the density profile and sphericity of halos,
we could derive $V_c$ easily as 
\begin{eqnarray}  \label{eq:V_c}
V_c^2(R) &=& \frac{GM(R)}{R}  \notag \\
&=& 4\pi G\beta\rho_cR^3_s\left[\frac{1}{R}\ln\left(1+\frac{R}{R_s}\right) - 
\frac{1}{R_s+R}\right]\ \ 
\end{eqnarray}
where $M(R)$ is the mass enclosed in radius $R$, and again this is
parameterized by $\beta$ and $R_s$. From a simulation point of view, it is
straightforward to measure $M(R)$ and then fit $\beta$ and $R_s$; from an
observational viewpoint, it is easy to measure $V_c(R)$, which could again
be used to fit $\beta$ and $R_s$.

To show how good the fittings could be, we pick out 9 halos with different
masses and sizes from our simulation box, fit the corresponding $\beta$ and $%
R_s$ using the measured $M(R)$, and plot these in Fig.~\ref{fig:profile_vc}.
As can be seen there, the fitting results (solid curves) agree with the
simulation results (crosses) quite well (in particular for the halo with $%
M=3.895\times10^{13}M_{\bigodot}$).

To see how this could be related to $\sqrt{\kappa}\varphi$, we remember that
in the above we have shown that $a\sqrt{\kappa}\varphi=4\zeta\Phi$, and so
all we need to do is to find an expression for $\Phi(R)$. For this, we use
the fact that the potential inside a spherical halo is given as 
\begin{eqnarray}
\Phi(R) &=& \int^{R}_{0}\frac{GM(r)}{r^2}dr + C
\end{eqnarray}
in which $GM(r)/r^2$ is the gravitational force and $C$ is a constant to be
fixed using the fact that $\Phi(R=\infty)=\Phi_\infty$ where $\Phi_\infty$
is the value of the potential far from the halo.

Using the formula for $GM(r)/r^2$ given in Eq.~(\ref{eq:V_c}) it is not
difficult to find that 
\begin{eqnarray}
\int^{R}_{0}\frac{GM(r)}{r^2}dr &=& 4\pi G\beta\rho_cR^3_s\left[\frac{1}{R_s}%
-\frac{\ln\left(1+\frac{R}{R_s}\right)}{R}\right]  \notag
\end{eqnarray}
and so 
\begin{eqnarray}
C &=& \Phi_\infty - 4\pi G\beta\rho_cR^2_s.
\end{eqnarray}
Then it follows that 
\begin{eqnarray}  \label{eq:Phi}
\Phi(R) &=& \Phi_\infty - 4\pi G\beta\rho_c\frac{R^3_s}{R}\ln\left(1+\frac{R%
}{R_s}\right).
\end{eqnarray}
If the halo is isolated, then $\Phi_\infty=0$ and we get 
\begin{eqnarray}  \label{eq:Phi_isolated_halo}
\Phi(R) &=& - 4\pi G\beta\rho_c\frac{R^3_s}{R}\ln\left(1+\frac{R}{R_s}%
\right).
\end{eqnarray}
However, in $N$-body simulations, we have a large number of dark matter
halos and no halo is ideally isolated from the others. In such situations, $%
\Phi_\infty$ in Eq.~(\ref{eq:Phi}) should be replaced by 
\begin{eqnarray}
\Phi_\ast\ \equiv\ \Phi(R=R_\ast\gg R_{\mathrm{vir}}) &\neq& 0
\end{eqnarray}
where $R_\ast$ is some radius large compared with $R_{\mathrm{vir}}$ (the
virialized halo radius) but small compared with inter-halo distances. Then
we get 
\begin{eqnarray}  \label{eq:delta_varphi}
a\sqrt{\kappa}\delta\varphi(R) &=& 4\zeta\left[\Phi_\ast - 4\pi G\beta\rho_c%
\frac{R^3_s}{R}\ln\left(1+\frac{R}{R_s}\right)\right].\ \ \ 
\end{eqnarray}

As an example to show how well Eq.~(\ref{eq:delta_varphi}) works, we show in
Fig.~\ref{fig:profile_scalfit} the results of $a\sqrt{\kappa}%
\delta\varphi(R) $ for the same halos used to fit $\beta$ and $R_s$ in Fig.~%
\ref{fig:profile_vc}. Here the crosses represent the values of $a\sqrt{\kappa%
}\delta\varphi$ measured from the $N$-body simulations and the curves our
analytical approximations, in which the solid curve is obtained by setting $%
\Phi_\ast=\Phi_\infty=0$ while the dashed curve is from tuning $\Phi_\ast$
appropriately. Obviously in most cases there is a (nearly) constant shift of
the approximation with respect to the numerical results, which accounts for
the nonzero-ness of $\Phi_\ast$.

Note that Eq.~(\ref{eq:delta_varphi}) captures the shapes for $a\sqrt{\kappa}%
\delta\varphi$ in various halos, but there is still one free parameter $%
\Phi_\ast$ to be tuned to match the numerical results. This parameter
summarizes our lack of knowledge about the environment which the considered
halo resides in. As a result, the formula Eq.~(\ref{eq:delta_varphi}) is
most suitable to apply in isolated halos, while for residential halos some
extra work remains to be done to make it accurate.

Alternatively, one could also consider Eq.~(\ref{eq:delta_varphi}) as a
3-parameter parameterization of $a\sqrt{\kappa}\delta\varphi$, for which the
three parameters $\beta, R_s$ and $\Phi_\ast$ could be fitted using the
results from $N$-body simulations. Given all the above results, we expect
that this could produce some nice fitting curves too, but we shall not
expand on this point here.

\section{Summary and Conclusion}

\label{sect:con}

To summarise, in this paper we have studied the behaviour of the \texttt{BSBM%
} varying-$\alpha$ model in the highly nonlinear regime of large scale
structure formation, with the aid of full $N$-body simulations that
explicitly solves the scalar field which controls the temporal and spatial
variations of $\alpha$.

We have checked that, because of the weak coupling to matter and the lack of
(nonlinear) potential, the scalar field is indeed very light everywhere and
thus does not cluster significantly, \emph{i.e.}, the spatial fluctuation of 
$\varphi$ is tiny. Because of this property, we have been able to simplify
the field equations, which in turn suggest that the scalar field
perturbation $\delta\varphi$ is proportional to the gravitational potential $%
\Phi$ [cf.~Eq.~(\ref{eq:ratio})]. The numerical simulations then conform
that such a simplification is justified to high precision.

We then concentrate on the profiles of scalar field inside virialized halos,
which galaxies (and observers) are supposed to reside in. Figs.~\ref%
{fig:profile_scal} and \ref{fig:profile_scal2} display the averaged profiles
of $a\sqrt{\kappa}\delta\varphi$ in the most massive halos from our
simulation box, and they show that $a\sqrt{\kappa}\delta\varphi$ decreases
as one goes from the halo centre outwards. In addition, the heavier the halo
is, the deeper the gravitational potential will be and, as a result of Eq.~(%
\ref{eq:ratio}), the larger $a\sqrt{\kappa}|\delta\varphi|$ is.
Interestingly, although $\delta\varphi$ does depends on the value of the 
\texttt{BSBM} parameter $\zeta$, the quantity $\delta\varphi/\varphi$ is
essentially independent of it (\emph{i.e.}, the same for \texttt{BSBM}
models with different $\zeta$ from our interested parameter space).

Thanks to the smallness of the scalar field coupling, the background
expansion rate and the source to the Poisson equation are essentially
unaffected by the scalar field, while at the same time the fifth force is
much weaker than ($\sim\zeta^2$ times) gravity so that its effect is also
negligible. As a result, the structure formation itself is indistinguishable
from that in pure $\Lambda$CDM model. In particular, the halo density
profiles are very well described by the NFW fitting formula. Furthermore,
the circular velocities $V_c$ inside halos from the simulations are also
well fitted using the analytical formula for $V_c$ derived assuming NFW
profiles [cf.~Fig.~\ref{fig:profile_vc}].

As another check of the fact that the scalar field perturbation $%
\delta\varphi$ is proportional to the gravitational potential $\Phi$
[cf.~Eq.~(\ref{eq:ratio})], we have derived an analytical expression for $a%
\sqrt{\kappa}\delta\varphi$ again by assuming NFW halo density profiles
[cf.~Eq.~(\ref{eq:delta_varphi})]. We adopt the NFW parameters fitted using
the circular velocities measured from simulation outputs in Eq.~(\ref%
{eq:delta_varphi}) and make predictions for $a\sqrt{\kappa}\delta\varphi$ in
different halos. As shown in Fig.~\ref{fig:profile_scalfit}, the predictions
agree very well with the numerical results for $a\sqrt{\kappa}\delta\varphi$%
, if we take into account the fact that halos are generally not isolated but
living in potential wells produced by other halos.

Our results suggest that for simple coupled scalar field models such as 
\texttt{BSBM} (and the one studied in \cite{Li:2010b}), the properties of
the scalar field perturbation could be studied without solving the scalar
field equation of motion explicitly (which is time-consuming). We could
either extract them from say the halo density profiles [using our Eq.~(\ref%
{eq:delta_varphi})] of $\Lambda$CDM $N$-body simulations, or from the data
on the galaxy rotation curves [using Eqs.~(\ref{eq:V_c}, \ref%
{eq:delta_varphi})] from observations. In the latter case, it is interesting
that two seemingly uncorrelated things could be studied together and only
once.

Let us stress that the above nice properties are only present because of the
smallness of the fluctuation in the scalar field, which is in turn due to
the lack of significant nonlinearity in its equation of motion. If the
scalar field is a chameleon, then its spatial fluctuation could be strong %
\citep{Li:2009, Li:2010, Zhao:2010} and it then becomes impossible to obtain
a simple analytical formula for $a\sqrt{\kappa }\delta \varphi $, such as
Eq.~(\ref{eq:delta_varphi}). $N$-body simulations will be the only tool to
study such models, which we hope to investigate in the future.

\bigskip

\section*{Acknowledgments}

The work described in this paper has been performed on \texttt{TITAN}, the
computing facilities at the University of Oslo in Norway, utilizing a
modified version of the \texttt{MLAPM} code \citep{Knebe:2001}.
Postprocessing is done on \texttt{COSMOS}, UK's National Cosmology
Supercomputer, and the halo properties are obtained using a modified version
of \texttt{MHF} \citep{mhf}. We thank George Efstathiou for encouragement
which helps bring this work to reality. BL is supported by the Research
Fellowship at Queens' College, Cambridge, and the Science and Technology
Facility Council (\texttt{STFC}) of the United Kingdom. DFM thanks the
Research Council of Norway FRINAT grant 197251/V30 and the Abel
extraordinary chair UCM-EEA-ABEL-03-2010. DFM is also partially supported by
the projects CERN/FP/109381/2009 and PTDC/FIS/102742/2008.

\end{document}